\documentclass[]{aa}  
\usepackage[]{amsmath}
\usepackage{graphicx}
\usepackage{amssymb}	
\usepackage{txfonts}
\usepackage[T1]{fontenc}
\usepackage[draft]{hyperref}
\usepackage{xcolor}
\begin{document} 

\title{Introducing a new multi-particle collision method for the evolution of dense stellar systems}

\subtitle{Crash-test $N$-body simulations}
\titlerunning{MPCDSS}
\author{Pierfrancesco Di Cintio\inst{1,2}, Mario Pasquato\inst{3,4,5}, Hyunwoo Kim\inst{6}, and Suk-Jin Yoon\inst{6}}
\institute{Dipartimento di Fisica e Astronomia \& CSDC, Universit\`a di Firenze, via G. Sansone 1, I--50019 Sesto Fiorentino, Italy\\
\email{pierfrancesco.dicintio@unifi.it}
\and
INFN - Sezione di Firenze, via G. Sansone 1, I--50019 Sesto Fiorentino, Italy
\and
INAF, Osservatorio Astronomico di Padova, vicolo dell'Osservatorio 5, I--35122 Padova, Italy
\and
Center for Astro, Particle and Planetary Physics (CAP$^3$), New York University Abu Dhabi\\
\email{mp5757@nyu.edu}
\and  
INFN- Sezione di Padova, Via Marzolo 8, I--35131 Padova, Italy
\and
Department of Astronomy \& Center for Galaxy Evolution Research, Yonsei University, Seoul 03722, Republic of Korea\\
\email{sjyoon0691@yonsei.ac.kr}\\}

\date{Received June 26, 2020; accepted November 18, 2020}
\abstract
{Stellar systems are broadly divided into collisional and non-collisional. 
While the latter are large-$N$ systems with long relaxation timescales and can be simulated disregarding two-body interactions, either computationally expensive direct $N$-body simulations or approximate schemes are required to properly model the former.
Large globular clusters and nuclear star clusters, with relaxation timescales of the order of a Hubble time, are small enough to display some collisional behaviour and big enough to be impossible to simulate with direct $N$-body codes and current hardware.}
{We introduce a new method to simulate collisional stellar systems, and validate it by comparison with direct $N$-body codes on small-N simulations.}
{The Multi-Particle Collision for Dense Stellar Systems ({\sc mpcdss}) code is a new code for evolving stellar systems with the multi-particle collision method. Such method amounts to a stochastic collision rule that allows to conserve exactly the energy and momentum over a cluster of particles experiencing the collision. The code complexity scales with $N \log N$ in the number of particles. Unlike Monte-Carlo codes, {\sc mpcdss} can easily model asymmetric, non-homogeneous, unrelaxed and rotating systems, while allowing us to follow the orbits of individual stars.}
{We evolve small ($N = 3.2 \times 10^4$) star clusters with {\sc mpcdss} and with the direct-summation code {\sc nbody6}, finding a similar evolution of key indicators. We then simulate different initial conditions in the $10^4 - 10^6$ star range.}
{{\sc mpcdss} bridges the gap between small, collisional systems that can be simulated with direct $N$-body codes and large non-collisional systems. {\sc mpcdss} in principle allows us to simulate globular clusters such as Omega Centauri and M54, and even nuclear star clusters, beyond the limits of current direct $N$-body codes in terms of the number of particles.}
\maketitle
\section{Introduction}
Our understanding of the formation and dynamical evolution of dense stellar systems such as globular clusters (hereafter GCs) and nuclear star clusters (hereafter NSCs) has a crucial impact on Galactic archaeology  (see e.g. \citealt{2018MNRAS.478..611B,2018MNRAS.478.5449M,2019MNRAS.488.1235M,2019A&A...630L...4M,2019NatAs...3..667I,2019A&A...632A...4D,2019ApJ...883L..31C,2020MNRAS.497.1603G}), multi-messenger astronomy \citep[where it allows us to better constrain compact object mergers; see e.g.][]{2002ApJ...572..407B, 2010MNRAS.402..371B, 2014MNRAS.440.2714B, 2014MNRAS.441.3703Z, 2016ApJ...830L..18B, 2016PhRvD..93h4029R, 2016PASA...33...36H, 2017MNRAS.464L..36A, 2017ApJ...836L..26C, 2018MNRAS.479.4652A, 2018PhRvL.120s1103K, 2019MNRAS.487.2947D, 2019ApJ...886...25B, 2019MNRAS.483.1233R, 2020MNRAS.tmp...31A}, cosmology and super-massive black hole science \citep[with star clusters acting both as nurseries of intermediate-mass black hole seeds and delivery mechanism to the galactic centers; see e.g.][]{1993ApJ...415..616C, 2001ApJ...562L..19E, 2004Natur.428..724P, 2006ApJ...641..319P, 2008MNRAS.388L..69C, 2012ApJ...750..111A,2014ApJ...796...40M, 2014MNRAS.444.3738A,2020arXiv200604922A}, and even stellar astrophysics \citep[as clusters are key to the formation of stellar exotica; see e.g.][]{1975MNRAS.172P..15F, 1995ARA&A..33..133B, 2001MNRAS.321..199P, 2004MNRAS.352....1F, 2006csxs.book..341V, 2007ApJ...661..210L, 2010ARA&A..48..431P, 2014ApJ...789...28P, 2019arXiv191007595V, 2020MNRAS.491..440W}.\\
\indent However, modelling self-gravitating $N$-body systems with a realistic number of stellar particles (in some cases well above $10^6$) over several relaxation times is extremely challenging in terms of computational resources due to the super-quadratic scaling of complexity with the number of particles in direct summation codes \citep[][]{1988ApJS...68..833M, 2003gnbs.book.....A}. Current state-of-the-art direct $N$-body simulations with $10^6$ particles need several months of computer time even on  dedicated accelerator Graphic Processing Units (GPUs) clusters to follow the evolution of typical globular clusters \citep[][]{2016MNRAS.458.1450W}.
This is an issue especially for simulating collisional systems, where the effects of relaxation driven by two-body interactions cannot be neglected.\\
\begin{figure}
    \centering
    \includegraphics[width=\columnwidth]{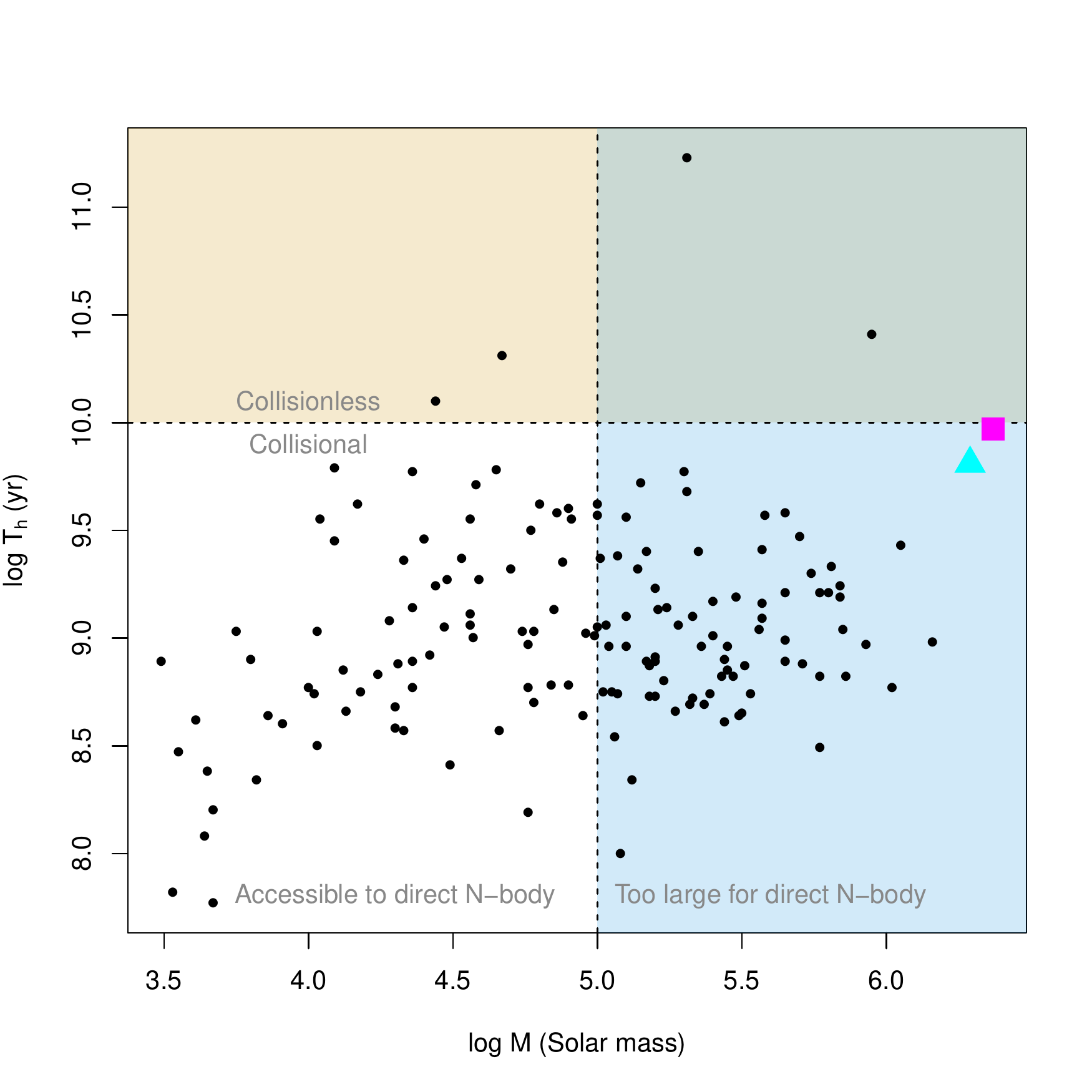}
    \caption{Globular star clusters of the Milky Way and satellites, from \cite{2005ApJS..161..304M}. Log half-mass relaxation time (in years) is plotted against log total mass (in Solar mass). Globular clusters Omega Centauri and M$54$ are shown as a magenta square and cyan triangle, respectively. Assuming a mean stellar mass of $0.5$ $M_\odot$, star clusters with mass above $10^5$ $M_\odot$ (shaded in light blue) contain $2 \times 10^5$ stars and can be simulated only with great computational effort and limited to a handful of realizations, especially if a realistic binary fraction is included, at variance of smaller systems (gray shaded area) that are well within the capabilities of current direct $N-$body codes. Star clusters with a relaxation time over the typical globular cluster age ($\approx 10$ Gyr) can be regarded as collisionless and are shaded in peach and dark green. According to this definition Omega Centauri and M$54$ are only slightly collisional, but clearly not accessible to modelling through direct $N$-body simulations. \label{GCs}}
\end{figure}
\indent In Fig.~\ref{GCs} we show that a large fraction of the Milky Way globular clusters is both in the collisional regime and contains a sufficiently high number of stars to make detailed modelling based on direct $N$-body simulations infeasible, especially when the need of obtaining a significant number of realizations of the same system is taken into account.\\
\indent Several approximated alternatives to the direct $N$-body approach\footnote{Note that direct $N$-body solutions are themselves not exact, due to the chaotic nature of the problem (see e.g. \citealt{2019MNRAS.489.5876D,2019arXiv191207406D} and references therein) and the finite precision of the numerics involved \citep[e.g. see ][which also proposes a creative alternative simulation scheme.]{2019arXiv191007291B}.} which do not share its prohibitive computational cost exist \cite[see][for an excellent review]{2016MmSAI..87..579H}.
The family of so-called Monte-Carlo codes, which essentially solve the Fokker-Planck equation \citep[][]{1971Ap&SS..13..284H, 1971Ap&SS..14..151H, 1975IAUS...69..133H, 1982AcA....32...63S, 1986AcA....36...19S, 2000ApJ...540..969J, 2001A&A...375..711F, 2001MNRAS.324..218G, 2002A&A...394..345F, 2006MNRAS.371..484G, 
2013ApJS..204...15P,
2013MNRAS.431.2184G, 2013MNRAS.429.1221H,2015MNRAS.453..605P, 2018ComAC...5....5R, 2019MNRAS.483.1523S} is perhaps the most successful among these. Monte-Carlo simulations however are generally limited to spherically symmetric systems, with the notable exception of the code developed by \cite{2015MNRAS.446.3150V}, which is unfortunately not in widespread use. Among other issues, this limitation was shown to lead to discrepancies between direct $N$-body and Monte-Carlo (or any Fokker-Planck solver that assumes spherical symmetry) in the presence of an external tidal field such as the Galactic one \citep[][]{2000ApJ...535..759T}, with the notable exception of the scheme by \cite{2014MNRAS.443.3513S} that shows a remarkably good agreement with direct $N-$body simulations with $N\approx 2\times 10^4$ when applied to GCs orbiting in a fixed point-like galactic potential.\\
\indent In this work, (the first of a series of three) we introduce a new simulation scheme, the  Multi-Particle Collision for Dense Stellar Systems (hereafter {\sc mpcdss}) code, which combines an essentially linear scaling of computational complexity in the number of particles with the ability to model configurations with arbitrary geometries. Here we introduce the structure of the code and present a first series of tests on the dynamical evolution of GCs, without including the stellar evolution modules, paving the way to the application to the two most massive star clusters in the Milky Way\footnote{With the exception of the NSC \citep[][]{2005ApJ...618..237W, 2011MNRAS.414.3699M, 2017IAUS..316...84N} which is in principle also amenable to simulation with {\sc mpcdss} and has been also studied with direct $N$-body simulations (\citealt{2011ApJ...729...35A,2014ApJ...784L..44P}) in the context of the so-called repeated accretion scenario (\citealt{2012ApJ...750..111A, 2014MNRAS.444.3738A}).}, M54 and Omega Centauri.
Both clusters show a spread in metallicity \citep[][]{1995AJ....109.1086S, 1999Natur.402...55L}, hinting at a non-trivial dynamical history which possibly includes one or more mergers \citep[see e.g.][]{2017MNRAS.466.2895C,2020IAUS..351...47A,2020ApJ...892...20A} which may still affect present-day observable properties \citep[][]{2013MNRAS.435..809A, 2016A&A...589A..95P}, and have large masses, thus being beyond the limit of what is currently modellable with ''honest" direct $N-$body simulations.\\
\indent This paper is structured as follows: in Section 2 we introduce the numerical methods used in {\sc mpcdss} to compute the gravitational field, treat the collisions and propagate the simulations particle trajectories, and we discuss the efficiency of our implementation. In Section 3 we compare a set of test simulations of collisional evaporation of smaller GCs using {\sc mpcdss} and the state-of-the-art direct $N-$body code {\sc nbody6} (\citealt{2003gnbs.book.....A,2012MNRAS.424..545N}). In Section 4 we present the results of numerical simulations of core collapse and mass segregation. In Section 5 we discuss our findings and, finally, Section 6 summarizes our results.
\section{Overview of the numerical method}
\subsection{Stochastic collisions: the Multi-Particle Collision (MPC) method}
In our numerical code we resolve the collisional interactions between stars using the so-called multi-particle collision method (hereafter MPC). originally, MPC was introduced by \cite{1999JChPh.110.8605M, 2004LNP...640..116M} in the context of numerical hydrodynamics for the simulation of mesoscopic fluids (e.g. polymers in solution, colloidal fluids). It has been shown that the method yields Galilean-invariant dynamics, that the Navier-Stokes equations are recovered in the continuum limit, and that relaxation towards thermodynamical equilibrium is correctly modeled \citep[see][for a detailed review]{2009acsa.book....1G}.
\begin{figure*}
    \centering
    \includegraphics[width=0.7\textwidth]{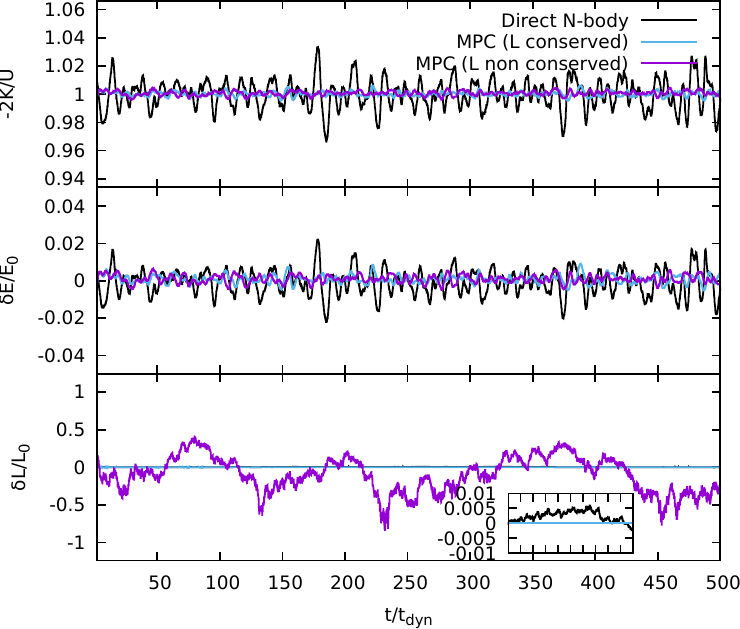}
    \caption{Evolution of the the virial ratio $-2K/U$ (top panel), the fluctuations of the total energy $\delta E$ in units of the initial energy $E_0$ (middle panel), and the norm of the total angular momentum $\delta L$ (bottom panel) for the same isotropic Plummer initial conditions with $N=2\times10^4$ particles, evolved with the direct $N-$body code (black lines) the exact angular momentum preserving MPC scheme (light blue lines) and the faster MPC rotation with random axis (purple lines). The initial conditions have a slight angular momentum due to the randomized initialisation procedure for stellar velocities.\label{virial}} 
\end{figure*}
\begin{figure*}
    \centering
    \includegraphics[width=0.9\textwidth]{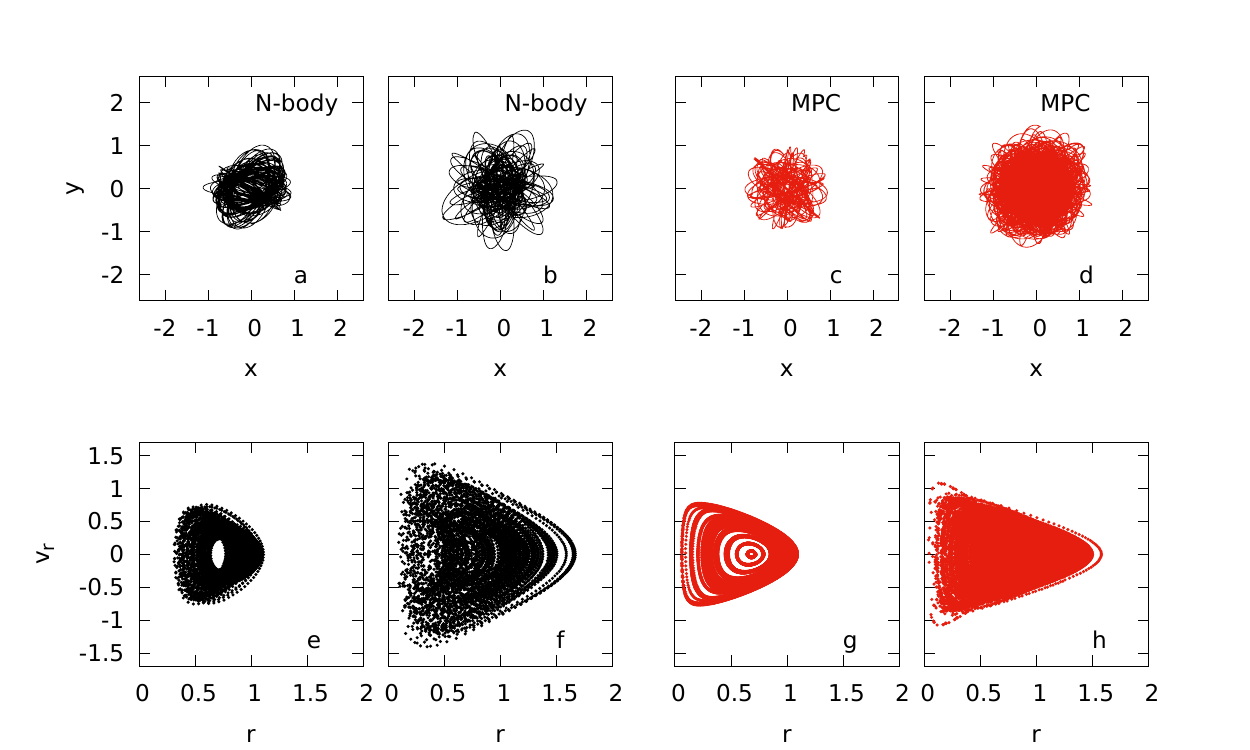}
    \caption{Projections on the $x-y$ plane of two tracer orbits starting from the same initial conditions propagated in a Plummer system with $N=2\times 10^4$ equal mass particles, evolved with an $N-$body code (panels a and b, black lines) and {\sc mpcdss} (panels c and d, red lines). The bottom panels (e,f,g,h) show the corresponding phase-space sections in the $r-v_r$ subspace.\label{orbits}} 
\end{figure*}
\begin{figure}
    \centering
    \includegraphics[width=0.95\columnwidth]{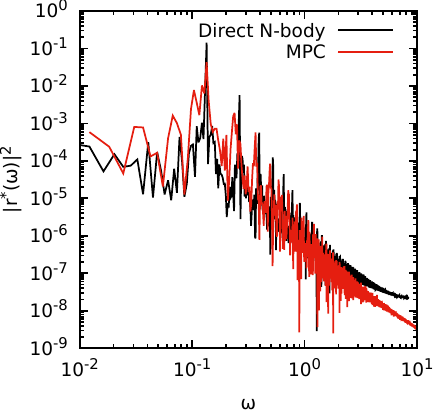}
    \caption{Fourier spectra of the radial coordinate $r$ for typical orbits extracted from an $N-$body simulation (red curve) and an MPC simulation (black curve) for a system of $N=2\times 10^4$ equal mass particles.\label{fourorbits}} 
\end{figure}
\begin{figure}
    \centering
    \includegraphics[width=0.95\columnwidth]{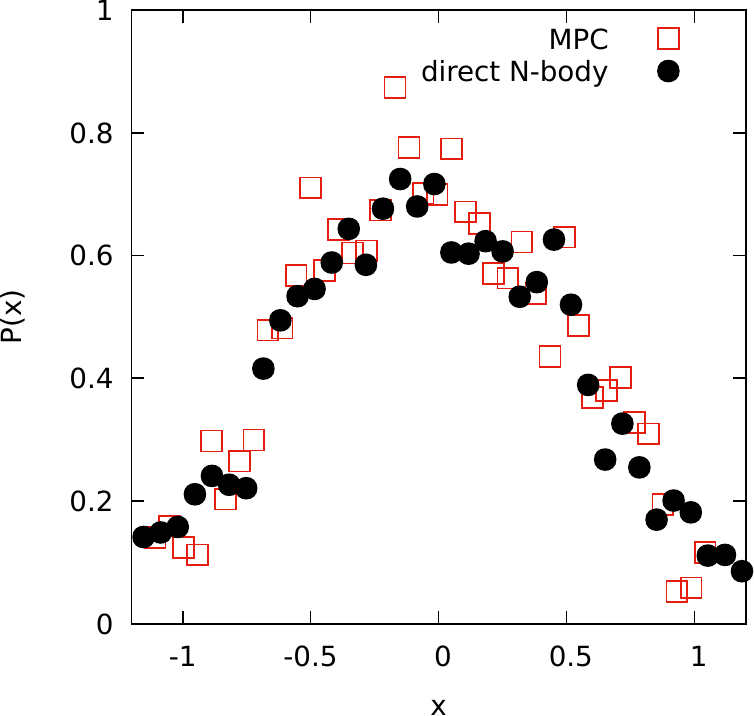}
    \caption{Probability density function of the $x$ coordinate $\mathcal{P}(x)$ for a tracer particle starting with the same initial condition in a $N=2\times 10^4$ Plummer model evolved with the direct $N-$body (black circles) and the MPC (red squares) methods.\label{pdf}}
\end{figure}
Recently, the MPC techniques have been also used in plasma physics to treat heat transport problems in reduced models in 1D (\citealt{2015PhRvE..92f2108D,2018CoPP...58..457C,10.1007/978-3-030-15096-9_10}) and 2D (\citealt{2017PhRvE..95d3203D,2017MmSAI..88..733D,2018MNRAS.475.1137D}).\\
\indent The MPC scheme alternates a streaming step (corresponding to non-collisional evolution) and a collision step. In three spatial dimension, the collision step amounts to a rotation of the particle's velocity vectors in the centre of mass frame of each cell\footnote{In the original implementation of the hybrid plasma PIC-MPC code, the cell structure is the same as the one used by the Maxwell solver routines to compute electromagnetic fields.} onto which the simulation domain has been partitioned.\\
\indent At the beginning of the collision step the code evaluates in every cell the centre of mass (c.o.m.) velocity
\begin{equation}
  \mathbf{u}_{\rm{com},i}=\frac{1}{m_{{\rm tot},i}}\sum_{j=1}^{n_i}m_j\mathbf{v}_j;\quad m_{{\rm tot},i}=\sum_{j=1}^{n_i}m_j
\end{equation}
and the relative velocities $\delta\mathbf{v}_j=\mathbf{v}_j-\mathbf{u}_i$. For each cell then, a random axis $\mathbf{R}_i$ and rotation angle $\alpha_i$ are sampled from uniform distributions. At this point, the vectors $\delta\mathbf{v}_j$ are rotated around $\mathbf{R}_i$ of $\alpha_i$ and then converted back to the simulation frame, so that for the $j-$th particle in cell $i$ the new velocity reads 
\begin{equation}\label{rotation}
\mathbf{v}_{j}^\prime=\mathbf{u}_i+\delta\mathbf{v}_{j,\perp}{\rm cos}(\alpha_i)+(\delta\mathbf{v}_{j,\perp}\times\mathbf{R}_i){\rm sin}(\alpha_i)+\delta\mathbf{v}_{j,\parallel},
\end{equation}  
where $\delta\mathbf{v}_{j,\perp}$ and $\delta\mathbf{v}_{j,\parallel}$ are the relative velocity components perpendicular and parallel to $\mathbf{R}_i$, respectively.\\
\indent Such operation conserves {\it exactly} the total kinetic energy $K_i$ and the three components of the momentum $\mathbf{P}_i$ in cell $i$ (e.g. see \citealt{tesiryder,2017PhRvE..95d3203D} for the rigorous proof). By introducing a constraint on the rotation angles $\alpha_i$, we conserve a component of the angular momentum vector of the cell $\mathbf{L}_i$ by defining $\alpha_i$ such that 
\begin{equation}\label{sincos}
{\rm sin}(\alpha_i)=-\frac{2a_ib_i}{a_i^2+b_i^2};\quad  {\rm cos}(\alpha_i)=\frac{a_i^2-b_i^2}{a_i^2+b_i^2},
\end{equation}
where
\begin{equation}\label{ab}
a_i=\sum_{j=1}^{N_i}\left[\mathbf{r}_j\times(\mathbf{v}_j-\mathbf{u}_i)\right]|_z;\quad b_i=\sum_{j=1}^{N_i}\mathbf{r}_j\cdot(\mathbf{v}_j-\mathbf{u}_i).
\end{equation}
In the formulae above, $\mathbf{r}_j$ are the particles position vectors, and the notation $[\mathbf{x}]|_z$ means that one is taking (without loss of generality) the component of the vector $\mathbf{x}$ parallel to the $z$ axis of the simulation's coordinate system, so that the $z$ component of the cell angular momentum is conserved.\\
\indent Note that, for strictly two dimensional systems, Equation (\ref{rotation}) becomes 
\begin{equation}
\mathbf{v}_{j}^\prime=\mathbf{u}_i+\mathcal{G}_{\alpha_i,i}\cdot\delta\mathbf{v}_{j},    
\end{equation}
where now $\mathcal{G}_{\alpha_i,i}$ is the 2D rotation matrix of an angle $\alpha_i$ that, if chosen according to Equations (\ref{sincos},\ref{ab}), ensures the conservation of the scalar angular momentum, (see \citealt{2017PhRvE..95d3203D}) in addition to $K_i$ and $\mathbf{P}_i$. Note also that, the conservation of the total angular momentum can be achieved even in three dimensional systems by choosing $\mathbf{R}_i$ to be parallel to the direction of the cell's angular momentum vector $\mathbf{L}_i$ and taking in the definition of $a_i$ the component of the vector $\left[\mathbf{r}_j\times(\mathbf{v}_j-\mathbf{u}_i)\right]$ parallel to the latter. For the simulations presented here, we limit ourselves to the standard rotation scheme with only one component of the total angular momentum conserved, as it is much less time consuming not having to determine cell by cell the direction of the angular momentum (pseudo)vector.\\ 
\indent As in GCs the collision frequency strongly depends on the local values of the stellar density and velocity dispersion, we condition the MPC step to a cell-dependent probability accounting for the local degree of collisionality. We define first the cell-dependent MPC probability as 
\begin{equation}\label{cumulative}
p_i={\rm Erf}\left(\beta\frac{\Delta t8\pi G^2\Bar{m}^2_i\Bar n\log\Lambda_i}{\sigma^3_i}\right),
\end{equation}
where $\Delta t$ is the timestep, $\Bar{n}$ the mean stellar number density, $\Bar{m}_i$ and $\sigma_i$ the average particle mass and the velocity dispersion in the cell, respectively and ${\rm Erf}(x)$ is the standard error function. The cell-dependent Coulomb logarithm is defined as 
\begin{equation}\label{logcoulomb}
\log\Lambda_i=\log(\sigma^2_ir_s/2G\Bar{m}_i),
\end{equation} 
with $r_s$ the typical scale length of the system. In the expression above $\beta$ is a dimensionless constant fixed to $2N_c$.\\
\indent Once Equation (\ref{cumulative}) is evaluated in each cell, a random number $p_{*i}$ is sampled from a uniform distribution in the interval $[0,1]$ and the multi-particle collision is applied for all cells for which $p_{*i}\leq p_i$. By doing so, particles in cells with higher collision frequency are more likely to be prone to a MPC step.\\
\indent Note that, applying a MPC scheme puts an additional constraint on the choice of the simulation time step $\Delta t$. In particular, if the latter is too short compared to a given cell crossing time (i.e. the time it takes to a test particle moving at the local mean speed to cross the cell) the collisions could be overestimated. Vice-versa, if $\Delta t$ is too large some fast particles might not spend sufficient time in cells at large collisionality (i.e. with local large density), virtually never experiencing collisions.
\subsection{Deterministic dynamics}
The collective dynamics of the systems is simulated by using a rather standard particle-mesh scheme (see e.g. \citealt{1988csup.book.....H}) that solves the Poisson equation 
\begin{equation}\label{poisson}
\Delta\Phi_\mathbf{r}=-4\pi G\rho_\mathbf{r}
\end{equation}
on a spherical grid in polar coordinates $N_r\times N_\vartheta\times N_\varphi$ and interpolates $\nabla\Phi$ at each particle position $\mathbf{r}_i$.\\
\indent The equations of motion between two MPC steps are solved with a standard second order leapfrog scheme with fixed timestep (see e.g. \citealt{mclachlan1992accuracy}) of the order of $10^{-2}$ in units of the system's initial crossing time $t_{\rm dyn}$ (see Equation \ref{tdyn} below). In the preliminary simulations presented in this work, in order to further speed up the calculations, instead of solving Equation (\ref{poisson}), we evaluate only the radial component of the gravitational field so that, in practice, the equations of motion become
\begin{equation}\label{mono}
\ddot{\mathbf{r}}_i=-\frac{GM(r_i)}{r_i^3}\mathbf{r}_i,
\end{equation}
where $M(r_i)$ is the mass within the particles radial coordinate $r_i$. By doing so, when needed, the potential $\Phi(r_i)$ on particle $i$ can be obtained as
\begin{equation}\label{pot}
\Phi(r_i)=-G\left(\frac{M(r_i)}{r_i}+\sum_{j=i+1}^N \frac{m_j}{r_j}\right),
\end{equation}
\begin{figure*}
    \centering
    \includegraphics[width=\textwidth]{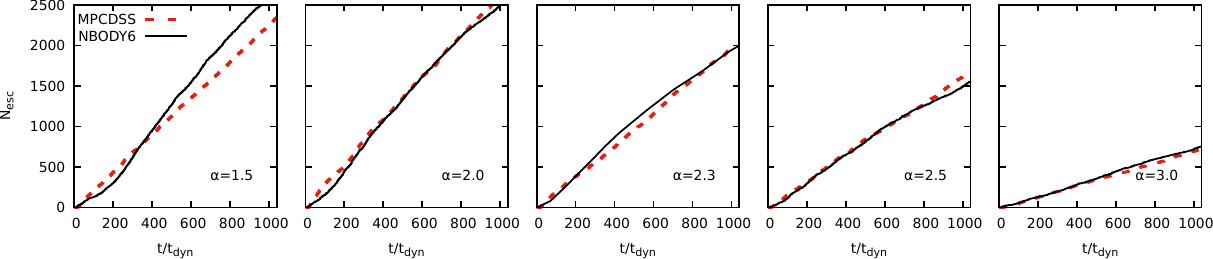}
    \caption{Number of escapers $N_{\rm esc}$ as a function of time in units of the dynamical time $t_{\rm dyn}$ for a Plummer model with $N=32000$ and power-law mass spectrum with, from left to right, $\alpha=1.5$, 2.0, 2.3, 2.5 and 3.0 when evolved with {\sc nbody6} (dashed lines) and with {\sc mpcdss} (solid lines).}\label{mpcvsnbody1}
\end{figure*}
\begin{figure*}
    \centering
    \includegraphics[width=0.7\textwidth]{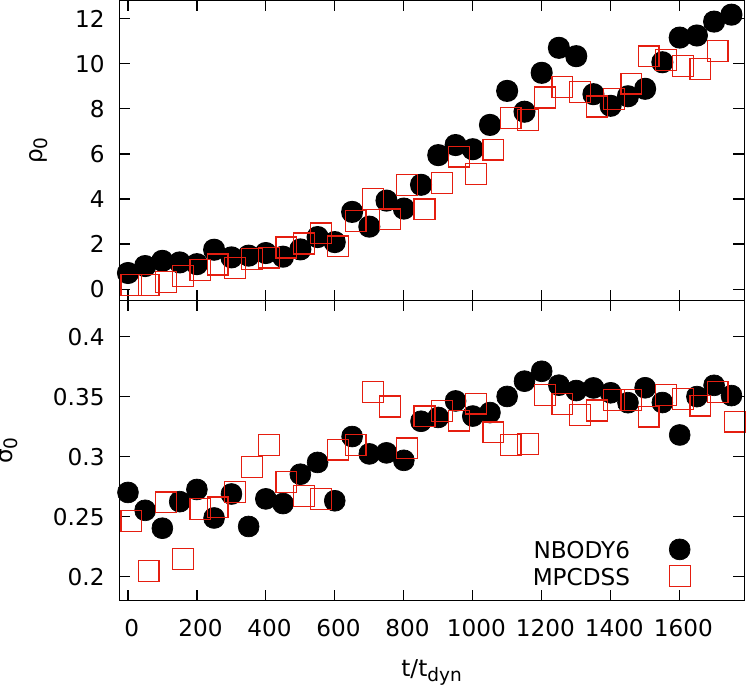}
    \caption{Evolution of the central density $\rho_0$ (upper panel) and central velocity dispersion $\sigma_0$ (lower panel) for the  $N-$body simulation (empty symbols) and the MPC simulation (filled symbols), for the model with $\alpha=2.3$ of Fig. \ref{mpcvsnbody1}. The symbol sizes are of the order of the mean (Poissonian) error of particle counts in both plots.\label{rho0}} 
\end{figure*}
\begin{figure*}
    \centering
    \includegraphics[width=0.6\textwidth]{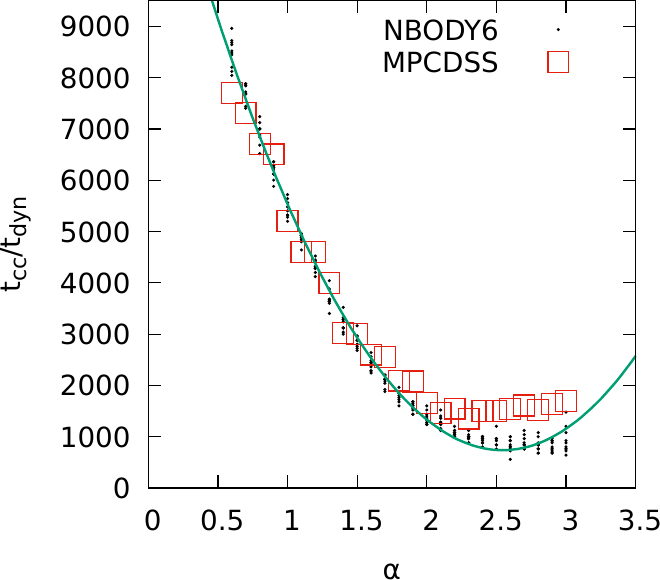}
    \caption{Time of core collapse $t_{cc}$ in units of the dynamical time $t_{\rm dyn}$ as function of the mass spectrum exponent $\alpha$ for MPC (red empty squares) and $N-$body (small black circles) simulations. A parabolic fit to the $N-$body simulations (green, thick line) is superimposed. The MPC result is generally within the range spanned by the direct $N$-body realizations, except for a few values of $\alpha$ on the high end, where core collapse happens earlier in direct $N$-body simulations.\label{tcorecoll}}
\end{figure*}
after having sorted the radial coordinates of all particles (see e.g. \citealt{2013ApJS..204...15P,2018ComAC...5....5R}) as in standard Monte-Carlo codes.\\
\indent With such assumptions the initial spherical symmetry of the model is preserved, as no radial orbit instability can take place (\citealt{2007cpms.conf..177C}) and, if the collision step is deactivated, each particle orbit retains its original plane and all angular momentum vectors $J_i=m_i\mathbf{r}_i\times\mathbf{v}_i$ are conserved individually.
\section{Comparison with direct $N$-body}
Introducing a somewhat stochastic mechanism of scattering between individual particle orbits via the MPC scheme, might at first seem to be inducing dramatic differences in the dynamical behaviour of orbit evolved with {\sc mpcdss} with respect to standard direct $N-$body integrators.
\subsection{Conservation laws and orbital structure}
As a first test of reliability of MPC simulations we have evolved small systems of $10000\leq N \leq 32000$ equal mass particles with the direct $N-$body code {\sc nbody6} and with {\sc mpcdss} and compared the evolution virial ratio $-2K/U$ (where $K$ is the total kinetic energy and $U$ the total potential energy), and the total energy and angular momentum fluctuations $\delta E=(E-E_0)$ and $\delta L=(L-L_0)$ (where $E_0$ and $L_0$ are the initial values of the energy and norm of the angular momentum, respectively) as resolved by the two schemes for identical isotropic equilibrium initial conditions. In addition, we have also studied the structure of the individual tracer particle orbits under both numerical methods.\\
\indent The models used for the two simulations sets have been set up as follows. We considered an isolated spherical isotropic model with \cite{1911MNRAS..71..460P} density distribution 
\begin{equation}\label{plummer}
\rho(r)=\frac{3}{4\pi}\frac{Mr_s^2}{(r_s^2+r^2)^{5/2}},
\end{equation}
with total mass $M$ and scale radius $r_s$, related to the half mass radius by $r_{\rm half}\approx 1.3r_s$. In both the $N$-body and MPC case the systems have been evolved up to $1000$ dynamical times defined by
\begin{equation}\label{tdyn}
t_{\rm dyn}=\sqrt{\frac{r_s^3}{GM}},
\end{equation}
with fixed timestep $\Delta t= t_{\rm dyn}/100$ and neglecting stellar evolution and formation of binaries (such that each simulation particle represents an individual star retaining its mass throughout the whole run). With such choice of $\Delta t$ we retain a good energy conservation while avoiding to overestimate the stochastic collisions in a given cell.\\
\indent As an example of the conservation of an equilibrium state, in Figure \ref{virial} we show the evolution of
the virial ratio as well as the total energy and angular momentum fluctuations $\delta E$ and $\delta L$ for an equilibrium isotropic \cite{1911MNRAS..71..460P} model with $2\times 10^4$ particles, evolved with the direct $N-$body code, and the two versions of {\sc mpcdss} with and without the angular momentum conserving scheme. We carry out the simulation only up to $500t_{\rm dyn}$ so that the collisions do not affect sensibly the model's properties. In all cases we have used the same fixed timestep $\Delta t$ and a second order propagation scheme.\\
\indent Remarkably, in the two MPC simulations, the virial equilibrium is preserved with smaller fluctuations with respect to the direct $N-$body integration. This is due to the fact that the enforced spherical symmetry and the smoother grid based potential in the {\sc mpcdss} simulations generally induce smaller force fluctuations on particles (for a given choice of $N$) with respect to a direct force calculation. Accordingly, the total energy shows only little fluctuations $\delta E/E_0$, as small as $10^{-3}$ for the MPC simulations at variance with the direct $N-$body simulations where $\delta E$ is already of the order of $10^{-2}$ at early times. When using the MPC scheme with enforced angular momentum conservation, for this particular choice of initial conditions and simulation set-up, its fluctuations $\delta L/L_0$ are of the order $10^{-4}$, while they are of order $10^{-3}$ for the direct $N-$body simulation (see the inset in lower panel of Fig.\ref{virial}). When the local conservation of angular momentum is not implemented in the MPC step, the total angular momentum might experience wild fluctuations $\delta L/L_0$ of the order of 0.5 the smaller the initial value of the angular momentum $L_0$. Remarkably, breaking the conservation of angular momentum does not affect sensibly the energy conservation nor the virial equilibrium.\\
\indent For what concerns the behaviour of individual particle orbits, surprisingly, the usage of the MPC step to resolve collisional processes does not alter qualitatively the behaviour of the orbits themselves with respect to standard $N-$body simulations. In the upper panels of Figure \ref{orbits} we show the projections on the $x-y$ plane of two orbits in a Plummer model with $N=2\times 10^4$ integrated with a direct $N-$body code (panels a and b, black lines) and with {\sc mpcdss} (panels c and d, red lines). Being always confined within less than two half-mass radii, the particles experience in both cases several ''close encounters" thus being subject to dramatic changes in orbital inclination and precession frequency. Direct $N-$body and MPC dynamics results in a large degree of phase-space ''diffusion" with particles exploring the whole energetically accessible region as shown in the lower panels of Fig. \ref{orbits}.\\
\indent Moreover, following \cite{2019MNRAS.489.5876D} we have also studied for a broad range of initial conditions the Fourier spectra of the radial coordinate $r$ for individual particle orbits. In general, the stochastic collision rule does not alter significantly the structure of the spectrum of a given orbit obtained from the same initial condition, with respect to a direct $N-$body evolution. In Figure \ref{fourorbits} we show the modulus squared $|r^*(\omega)|^2$ of the radial coordinate $r$ for a particle propagated in the same Plummer model with $N=2\times 10^4$ with the two simulation approaches, corresponding to panels a and c of Fig. \ref{orbits}. Remarkably the structure of the fundamental frequency (and a large fraction of higher harmonics at larger values of $\omega$) are preserved, thus leading to speculate that MPC dynamics can be sufficiently trusted even for larger systems, as the introduction of a stochastization of particles velocities does affect the collective behaviour of the models (even for a rather small $N$ sucha as $2\times 10^4$). In addition, for the same orbits of Figure \ref{fourorbits}, in Fig. \ref{pdf} we show the probability density function of their $x$ coordinate $\mathcal{P}(x)$ (i.e. the distribution of positions along $x$ attained by the tracer particles within a given time interval) up to $t=1000t_{dyn}$. The two distributions match remarkably well, thus confirming the overall agreement between the individual orbital structure in the two simulation methods. We note that, a similar comparison to $N-$body orbits has been made also for massive tracer particle orbits propagated semi-analitically with Fokker-Planck schemes (see e.g. \citealt{2002PhRvL..88l1103C,2002ApJ...572..371C,2003ApJ...592...32C}) or direct integration of the Langevin equation (\citealt{2020IAUS..351...93D}) in $N=10^5$ equal mass Plummer models. In general, at least for the models considered in these studies, the distribution of the coordinates $\mathcal{P}(x)$ evaluated from the numerical solution of the Fokker-Planck equation has the best agreement with the results of the direct $N-$body simulations, with the Langevin approach being strongly dependent of the form of the noise term. Our MPC approach places itself in between the two methods.\\
\indent We note that, using the MPC scheme results in a dramatic reduction of the computational cost of the simulation while retaining sufficiently reliable results. For example, the evolution of an $N=32000$ cluster up to $2000t_{\rm dyn}$ takes a few days on a dedicated GPU workstation for the direct $N-$body simulation without stellar evolution and adaptive time step. When using a $N_c=32\times 16\times 16$ grid in polar coordinates to perform the collisions and an $N_r=2000$ radial grid to evaluate the gravitational field in monopole approximation (see Eq. \ref{mono}), our MPC simulations with fixed $\Delta t$ take roughly one hour on a single core of a 64 bit {\sc INTEL\textregistered} machine.
\subsection{Core collapse with mass spectrum}
The effect of multiple mass populations on the dynamical evolution of GCs is of prime importance. The present implementation of our MPC code offers the interesting chance to study multi-mass systems without adding extra computational complexity. We performed a set of additional tests with {\sc nbody6} and with {\sc mpcdss} simulating the evolution up to and after core collapse of Plummer models with mass spectrum.\\ 
\indent For the sake of simplicity, instead of using the multi-slope \cite{2002Sci...295...82K} mass function, in this work the particle masses $m_i$ have been extracted from a pure power-law mass function of the form 
\begin{equation}\label{mspectrum}
\mathcal{F}(m) = \frac{C}{m^{\alpha}};\quad m_{\rm min} \leq m\leq m_{\rm max}, 
\end{equation}
where $\alpha>0$, and the normalization constant $C$ depends on the minimum-to-maximum-mass ratio $\mathcal{R}=m_{\rm min}/m_{\rm max}$ so that $\int_{m_{\rm min}}^{m_{\rm max}}\mathcal{F}(m){\rm d}m=M$.\\
\indent We run simulations with $N=32000$ for a range of $\alpha$ spanning from $0.6$ to $3.0$ by intervals of $0.1$, and in the case of the direct $N$-body simulations we run $10$ different realizations (with a different seed for the initial conditions) for each value of $\alpha$. In all cases with mass spectrum discussed in this work we fix $\mathcal{R}$ to $10^{-3}$.\\
\indent In both the $N$-body and MPC case the systems have been evolved for $10^4$ dynamical times, corresponding to roughly 20 two-body (collisional) relaxation times of a model with the same total mass and number of equal mass particles, given by
\begin{equation}\label{t2body}
t_{2b}\approx\frac{0.138N}{\log N}t_{\rm dyn}.  
\end{equation}
\begin{figure*}
    \centering
    \includegraphics[width=0.8\textwidth]{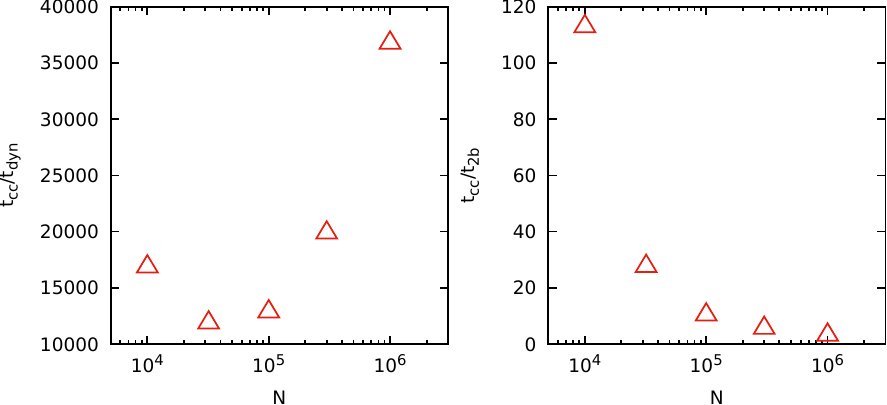}
    \caption{Relations of the number of particles $N$ versus the time of core collapse $t_{cc}$ in units of the dynamical time (left panel) and
relaxation time (right panel) for single component Plummer models}\label{tcc}
\end{figure*}
\begin{figure}
    \centering
    \includegraphics[width=\columnwidth]{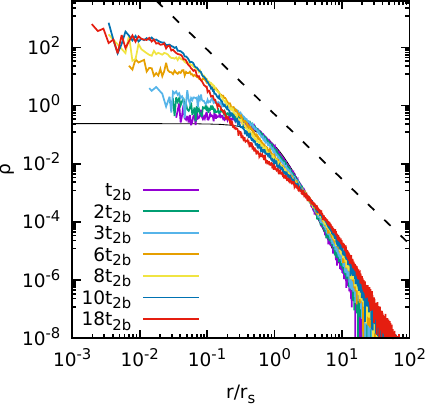}
    \caption{3D density profile for a model with $N=10^5$ equal mass particles and initial Plummer density distribution (thin black solid line) at $t=1,$ 2, 3, 6, 8, 10 and 18 two-body relaxation times $t_{2b}$ (coloured lines). The heavy dashed line marks the theoretical $r^{-2.23}$ profile.\label{corecollu}}
\end{figure}
\begin{figure}
    \centering
    \includegraphics[width=0.9\columnwidth]{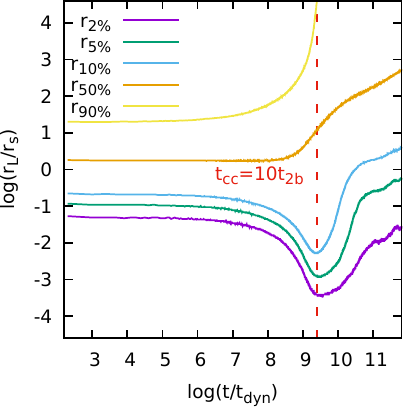}
    \caption{Evolution of the 3D Lagrangian radii enclosing $2\%$, $5\%$, $10\%$, $50\%$ and $90\%$ of the total number of particles $N$ for the same model in Fig. \ref{corecollu}. The vertical dashed line marks the system's core collapse time $t_{cc}\approx10t_{2b}\approx12000t_{\rm dyn}$\label{lagrangian}}
\end{figure}
Again, as a rule, in all sets of {\sc mpcdss} simulations we have fixed the timestep $\Delta t$ to $t_{dyn}/100$, neglected stellar evolution and the contribution of binaries\footnote{Note that in direct $N$-body simulations, even if started without binaries, may form a few new binaries dynamically.}.\\
\indent For the two simulations sets ($N-$body and MPC) we have extracted and compared as a function of time the number of escapers $N_{\rm esc}$ (defined as the number of particles being at $r>17r_s$ with positive energy, see e.g. \citealt{2000MNRAS.318..753F}), the central density and velocity dispersion $\rho_0$ and $\sigma_0$, as well as the mass function in the core.\\
\indent In Figure \ref{mpcvsnbody1} we present the time evolution of the number of escapers $N_{\rm esc}$ for choices of $\alpha$ in Equation (\ref{mspectrum}). We find that in all cases the MPC evolution (solid line) recovers the quasi-linear trend of $N_{\rm esc}$ with time. However, some discrepancies between MPC and $N-$body simulations are observed in particular at low values of $\alpha$ (i.e. ''flatter" mass spectrum). We interpret such difference as the effect of strong two-body collisions between light and heavy particles, resolved in a direct $N-$body code but somewhat smeared-out in a multi-particle collision code. In practice, in models with a mass spectrum with a larger fraction of heavy particles, it is more likely that the lighter ones are kicked out with positive energy when experiencing close encounters with heavier stars as compared with systems with mass spectra strongly peaked at low masses. Since the MPC step, even in presence of a mass spectrum, simulates multiple inter-particle collisions with a single move acting on all particles of the cell, the contribution of few but strong light-heavy particle encounters is obviously underestimated.\\
\indent For the best agreement case, represented by the simulations with $\alpha=2.3$ (and corresponding to a \citealt{salpeter55} mass function), we show in Figure \ref{rho0} the evolution of the central density $\rho_0$ and central velocity dispersion $\sigma_0$ defined within the Lagrangian radius enclosing 8\% of the total mass $M$. In this case, the evolution of both quantities with the MPC code (squares) matches remarkably well that obtained using the {\sc nbody6} (circles).\\
\indent In addition, for each simulation of the two sets we take the time of core-collapse $t_{cc}$ as the time at which the minimum value reached by $r_{2\%}$ (i.e. the 3D Lagrangian radius enclosing the 2\% of the total mass). Fig.~\ref{tcorecoll} shows the time of core collapse as a function of the initial mass-function power-law exponent $\alpha$. As expected, simulations starting with a shallower mass-function have heavier particles, slowing down the core collapse. Thus low-$\alpha$ runs take longer to reach core-collapse. Interestingly, for large values of $\alpha$ (say $\gtrsim 2.6$) $t_{cc}$ starts increasing again in both MPC and direct $N-$body simulations. We speculate that such curious behaviour might be due to the less efficient dynamical friction in models dominated by low mass particles (see \citealt{2010AIPC.1242..117C}), resulting in larger sinking time scales for particles sitting in the large mass part of the spectrum. To guide the eye, we fit the relation between $\alpha$ and $t_{cc}$ using a second order polynomial (green solid line). We performed additional simulations (to be published elsewhere) with different values of $N$, $\alpha$ and $\mathcal{R}$ and the non-monotonic trend of $t_{cc}$ with $\alpha$ appeared to be robust. Interestingly, the time of core collapse estimated with MPC simulations appears to be systematically larger that its counterpart for direct $N-$body simulations for values of $\alpha$ larger than 2.3. We speculate that the reason of this discrepancy lies in the absence of a binary formation mechanism in the implementation of {\sc mpcdsse} used here. This, combined with the less efficient dynamical friction results in the time at which the collapse inverts being delayed, thus yielding a larger $t_{cc}$.
\begin{figure*}
    \centering
    \includegraphics[width=0.9\textwidth]{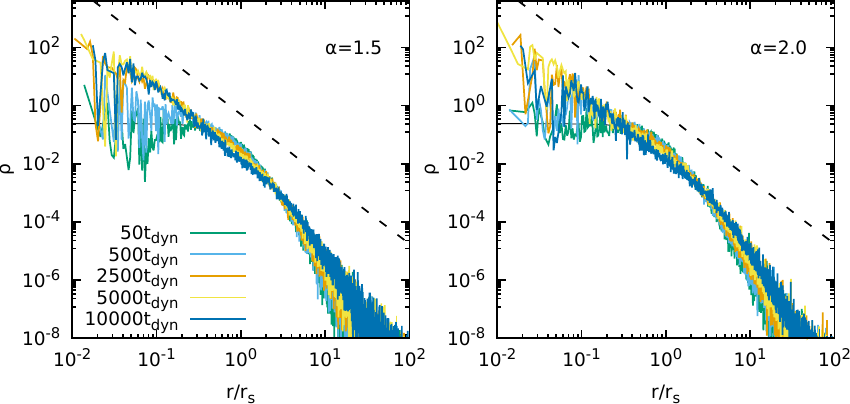}
    \caption{Evolution of the 3D density profile for two models with $N=10^5$ and initial Plummer density distribution (thin solid line) and mass spectra with $\alpha=1.5$ (left panel) and $2.0$ (right panel). As in Fig. \ref{corecollu}, the heavy dashed line marks the theoretical $r^{-2.23}$ profile.\label{alpharho}}
\end{figure*}
\section{Further numerical experiments and results}\label{results}
\begin{figure}
    \centering
    \includegraphics[width=0.9\columnwidth]{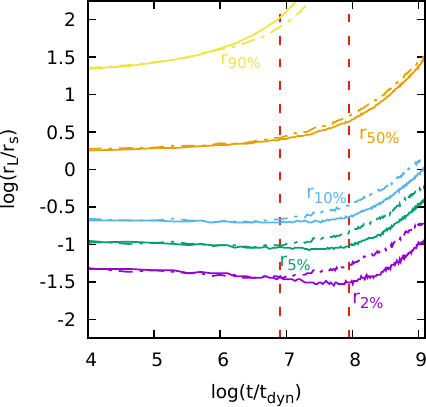}
    \caption{Evolution of the 3D Lagrangian radii enclosing $2\%$, $5\%$, $10\%$, $50\%$ and $90\%$ of the total number of particles $N=10^5$ for the two models shown in Fig. \ref{alpharho}. The solid and dotted-dashed lines refer the $\alpha=1.5$ and 2 cases, respectively. The two vertical dashed lines mark the core collapse times for the models $t_{cc}\approx 2000t_{\rm dyn}$ for $\alpha=1.5$ and $t_{cc}\approx 773t_{\rm dyn}$ for  $\alpha=2.0$.\label{lagalpha}}
\end{figure}

We have performed a broad spectrum of numerical experiments to determine the range of applicability of our newly introduced simulation method. First of all we have investigated the process of core collapse of single component models starting with isotropic Plummer density profiles in a broad range of systems sizes spanning from $10^3$ to $10^6$. For this set of numerical experiments we follow the evolution of the three dimensional and the associated Lagrangian radii containing different fractions of the total number of particles (or mass) between 2\% and 90\%.\\
\indent In line with expectations, we find from the MPC simulations that for large values of $N$ the time of core collapse becomes asymptotically larger (in units of the dynamical time $t_{\rm dyn}$), as shown in the left panel of Figure \ref{tcc} for the equal masses case with $N$ ranging from $10^4$ to $10^6$. Remarkably, for low values of $N$ there seems to be a somewhat non-monotonic trend of $t_{cc}$ with $N$. 
We performed additional MPC simulations with $N$ as small as $10^4$ using different realizations of the initial condition with choices of timestep and grid size, and such trend persists.\\
\indent When expressing $t_{cc}$ in units of the collisional relaxation time $t_{2b}$ (cfr. Eq. \ref{t2body}), the picture is inverted and large $N$ systems reach core collapse earlier in units of their intrinsic $t_{2b}$, see right panel of same Figure. As $N$ increases to larger and larger values, the systems start to approach the so-called collisionless limit, where the effects of collisions become more and more negligible, and therefore the process of core collapse is driven mainly by collective instabilities that take place on increasingly large timescales. On the other hand, as $N$ increases, $t_{2b}$ also increases as $N/\log{N}$ resulting in a {\it decreasing} trend of $t_{cc}/t_{2b}$ with $N$.\\ 
\indent In Figure \ref{corecollu} we show for the $N=10^5$ case the radial density profile at different times between $1t_{2b}$ and $18t_{2b}$. It is evident as at already around $2t_{2b}$ (corresponding roughly to $2340t_{\rm dyn}$ for this value of $N$) the density has significantly departed from the initial Plummer profile (marked in figure by the thin black line). Remarkably, at later times the inner part of the density slope approaches the $r^{-2.23}$ trend (heavy dashed line) as predicted by \cite{cohn80} (see also \citealt{1988MNRAS.230..223H}), and in nice agreement with the Monte-Carlo simulations by \cite{2000ApJ...540..969J} and \cite{2013ApJS..204...15P}. For times larger than roughly $8t_{2b}$ the re-expansion of the outer regions becomes evident, as can be seen also form the evolution of the Lagrangian radii in Figure \ref{lagrangian}. Surprisingly, even if in the current implementation of {\sc mpcdss} there is no explicit treatment of the binaries, the evolution of the density profile towards and beyond the core collapse matches remarkably well with its counterpart obtained with different numerical methods incorporating a systematic treatment of binaries. In particular, the fact that the critical $r^{-2.23}$ density slope is recovered in MPC simulations (even for systems as small as $N=10^4$ and as large as $N=10^6$, not shown here) suggests that it is not strongly related to the ''thermostat effect" of binary formation, but rather to the evaporation of particles  from the contracting core.\\
\indent We find a core-collapse time $t_{cc}$ (i.e. the time at which the central part of the cluster reaches the highest concentration that we measure here as the minimum attained by the Lagrangian radius containing the $2\%$ of the simulation particles) of about $10t_{2b}$ (indicated by the vertical line in Fig. \ref{lagrangian}), in rather good agreement with the Monte-Carlo simulations of \cite{2000ApJ...540..969J,2012MNRAS.425.2872H} and the $N-$body simulations of \cite{2008MNRAS.389..889K}, finding values between $10$ and $15t_{2b}$ for models with initial conditions analogous to the ones used in our simulations (i.e. Plummer profile, $N=10^5$ and no mass spectrum).\\
\indent Moreover, we have also performed additional simulations for Plummer models with different values of $N$ and mass spectra finding that, surprisingly, for the models with $\alpha$ in the range between 1.5 and 3 the asymptotic slope of the density profile in the inner regions has a better matching to the predicted $r^{-2.23}$ trend, as shown in Fig. \ref{alpharho} for the $\alpha=1.5$ and $\alpha=2$ cases with $N=10^5$.\\
\indent We observe that, in general, for fixed number of particles and total mass $M$, models with a mass spectra reach the core collapse faster in units of $t_{\rm dyn}$ than the associated equal mass case. This can be seen from Figure \ref{lagalpha} where we show  the Lagrangian radii enclosing the same fractions of the total number of particles as in Fig. \ref{lagrangian}, but for two models with mass spectrum with exponents $\alpha=1.5$ and 2. In both cases the core collapse time $t_{cc}$ is well below $10^3t_{\rm dyn}$ being roughly $2000t_{\rm dyn}$ for $\alpha=1.5$ and $773t_{\rm dyn}$ for  $\alpha=2.0$ as marked by the vertical lines in figure.\\
\indent Since the time dependent radii enclosing a given fraction of the total mass $M$ or number of particles $N$ are not the same quantity for a model with different species, we have evaluated both types of ''Lagrangian radii" for some of the models with mass spectrum. As expected, since as result of the more efficient dynamical friction, more massive stars tend to accumulate to the centre of the system, the Lagrangian radii computed for a given percentage of the mass of the model attain systematically smaller values than those evaluated for the same percentage of the total number of particles instead. However, the estimated core collapse times do not differ significantly with the two choices, independently on the number of particles in the simulation.     
\section{Discussion, conclusions and future prospects}
We have introduced a new code for simulating the collisional evolution of dense stellar systems using the Multi Particle Collision (MPC) approach. Our code is characterized by an $N \log N$ scaling in the number of stars, which makes it suitable for simulating massive GCs and the Milky Way NSC. These systems are for the foreseeable future well beyond the reach of direct-summation $N$-body codes because the latter scale quadratically with the number of stars. 
The MPC method is based on alternating streaming steps (where stars evolve in the smooth gravitational potential of the whole star system) and collision steps which are meant to model the relaxation effects induced by stellar encounters. Collision steps are cell-based rotations of the stellar velocity vectors which by construction conserve mass, momentum, and energy\footnote{In our implementation of the method, angular momentum conservation is also insured, but this feature can be switched off to speed up calculations if necessary.}. The MPC approach abstracts away from the intricacies of two- and multiple-body encounters that require in direct $N$-body codes techniques such as softening or Kustaanheimo-Stiefel regularization (see e.g. \citealt{2008LNP...760...31M}) in order to treat the singularity of the $1/r$ gravitational potential\footnote{ Note that, adaptively softening the gravitational interaction with $1/\sqrt{\epsilon^2+r^2}$ or preforming the Kustaanheimo-Stiefel mapping of position $r$ and time $t$ in  $u=\sqrt{r}$ and ${\rm d}t=r{\rm d}\tau$ every time a two body close encounter takes place, adds further increase in computational time in direct simulations.}, while retaining the relaxation effects of the encounters. Doing away with the need of computing all pairwise forces, the MPC approach results in much lower algorithmic complexity with the number of stars, without losing the ability to correctly recover the long term evolution of stellar systems.\\
\indent Compared to Monte-Carlo approaches, our code has the advantage of easily simulating any geometry as the MPC method has no sensible dependence on the shape of the individual cells or the overall grid structure, while most Monte-Carlo codes are confined to highly symmetric configurations. We can thus simulate rotating\footnote{Note that, since the MPC operator acts in each cell's center of mass frame, the presence of a collective rotation field does not affect in any way the stochastic collision mechanism.}, merging or tidally disrupted star clusters with no additional effort and no loss in accuracy with respect to spherically symmetric systems.\\
\indent In this paper presented our MPC code and run a few test simulations, showing that the total energy and angular momentum of an isolated simulation are conserved. We also calculated the virial ratio (kinetic over potential energy), which is also conserved with remarkable accuracy.
Finally, we validated our code by comparison to direct $N$-body simulations of star clusters. We find that the evolution of the central density, central velocity dispersion, and number of escapers as a function of time in MPC simulations closely follows that in direct $N$-body simulations over a wide a range of stellar mass spectra. Additionally, the time at which core collapse is reached is also in good agreement with both direct $N$-body and theoretical analytical calculations.\\
\indent In the future we plan to add an interface with community stellar evolution module such as Single Stellar Evolution \citep[SSE; ][]{2000MNRAS.315..543H} or the more recent SEVN \citep{2017MNRAS.470.4739S} and to introduce one or more schemes to simulate binary stars, such as for example tracked particles with internal dynamics in the same spirit of plasma codes including ionization and recombination (see e.g. \citealt{2013PhRvL.111l3401D}), or a nested direct integrator for a subset of particles \citep[e.g. ][]{2012ascl.soft08011F} coupled with a binary evolution code \citep[][]{2002MNRAS.329..897H, 2018MNRAS.474.2959G} in order to study the dynamics of compact objects \citep[e.g.][]{2016MNRAS.459.3432M,2020arXiv200302277R} and other phenomena hinging critically on binary modeling, such as blue stragglers \citep[][]{2015ApJ...799...44M, 2018ApJ...867..163P, 2020A&A...640A..79P}. We will then run a large set of simulations which we plan to release publicly along with the fully parallelized version of the code, including simulations meant to model specific objects, such as Omega Centauri and M~$54$. Finally, we will address the black-hole retention problem in star clusters, focusing in particular on the fate of intermediate-mass black holes in the Galactic nuclear cluster.
\begin{acknowledgements}
This project has received funding from the European Union's Horizon $2020$
research and innovation program under the Marie Sk\l{}odowska-Curie grant agreement No. $664931$. This material is based upon work supported by Tamkeen under the NYU Abu Dhabi Research Institute grant CAP3.
P.F.D.C. wishes to thank the financing from MIUR-PRIN2017 project \textit{Coarse-grained description for non-equilibrium systems and transport phenomena
(CO-NEST)} n.201798CZL. S.-J.Y. acknowledges support by the Mid-career Researcher Program (No.2019R1A2C3006242) and the SRC Program (the Center for Galaxy Evolution
Research; No. 2017R1A5A1070354) through the National Research
Foundation of Korea. We thank A.A. Trani, L. Ciotti and G. Ciraolo for the discussions at an early stage of this project, and the anonymous Referee for his/her comments that helped improving the presentation of this work.  
\end{acknowledgements}
\bibliographystyle{aa}
\bibliography{ms}
\end{document}